\documentclass[prd,twocolumn,tightenlines,nofootinbib,superscriptaddress]{revtex4-1}
\usepackage{bm,dcolumn,amsmath,graphicx,amsfonts,amssymb}

\usepackage{hyperref} 
\usepackage{xcolor}
\definecolor{cite}{rgb}{0.,0.,0.85}   
\hypersetup{colorlinks,linkcolor={cite},citecolor={cite},urlcolor={cite}}

\newcommand{\bra}[1]{\ensuremath{\langle #1|}}	
\newcommand{\ket}[1]{\ensuremath{|#1\rangle}}	
\newcommand{\braket}[1]{\ensuremath{\langle #1\rangle}}	

\newcommand{\threej}[6]{\ensuremath{\begin{pmatrix}#1&#2&#3\\#4&#5&#6\end{pmatrix}}}	
\renewcommand{\v}[1]{\ensuremath{\boldsymbol{#1}}}		
\newcommand{\E}[1]{\ensuremath{\times10^{#1}}}	
\newcommand{\twocomp}[2]{\ensuremath{\begin{pmatrix}#1\\#2\end{pmatrix}}}	
\newcommand{\vhat}[1]{\ensuremath{\hat{\boldsymbol{#1}}}}		

\newcommand{\be}{\begin{equation}}
\newcommand{\ee}{\end{equation}}

\def\d{\ensuremath{{\rm d}}}

\def\h{\ensuremath{\hbar}} 
 
\def\en{\ensuremath{\varepsilon}}

\renewcommand{\min}{\text{min}}

\renewcommand{\a}{\ensuremath{\alpha}} 
\renewcommand{\b}{\ensuremath{\beta}}
 
\newcommand{\s}{\ensuremath{\sigma}}

\newcommand{\w}{\ensuremath{\omega}} 

\newcommand{\un}[1]{\ensuremath{\,{\rm{#1}}}} 

\usepackage[normalem]{ulem}
\definecolor{newc}{rgb}{0.,0.6,0.4}

\def\sbe{\ensuremath{\bar\sigma_e}}

\begin{document} 

\title{Electron-interacting dark matter:\ Implications from DAMA/LIBRA-phase2\\ and prospects for liquid xenon detectors and NaI detectors}

\author{B.\ M.\ Roberts}\email[]{b.roberts@uq.edu.au}
	\affiliation{SYRTE, Observatoire de Paris, Universit\'e PSL, CNRS, Sorbonne Universit\'e, LNE, 61 avenue de l'Observatoire 75014 Paris, France}
		\affiliation{School of Mathematics and Physics, The University of Queensland, Brisbane, QLD 4072, Australia}
\author{V.\ V.\ Flambaum}
	\affiliation{School of Physics, University of New South Wales, Sydney 2052, Australia}
	\affiliation{Johannes Gutenberg-Universit\"at Mainz, 55099 Mainz, Germany}
\date{ \today }  

\begin{abstract}
We investigate the possibility for the direct detection of low mass (GeV scale) WIMP dark matter in scintillation experiments.
Such WIMPs are typically too light to leave appreciable nuclear recoils, but may be detected via their scattering off atomic electrons.
In particular, the DAMA Collaboration  [R.~Bernabei {\sl et al.}, Nucl.~Phys.~At.~Energy {\bf19}, 307 (2018)] has recently presented strong evidence of an annual modulation in the scintillation rate observed at energies as low as 1\,keV.
Despite a strong enhancement in the calculated event rate at low energies, we find that an interpretation in terms of electron-interacting WIMPs cannot be consistent with existing constraints.
We also demonstrate the importance of correct treatment of the atomic wavefunctions, and show the resulting event rate is very sensitive to the low-energy performance of the detectors, meaning it is crucial that the detector uncertainties be taken into account.
Finally, we demonstrate that the potential scintillation event rate can be much larger than may otherwise be expected, meaning that competitive searches can be performed for $m_\chi\sim$\,GeV scale WIMPs using the conventional prompt ($S1$) scintillation signals.
This is important given the recent and upcoming very large liquid xenon detectors.
\end{abstract} 

\maketitle

\section{Introduction}

The identity and nature of dark matter (DM) remains one of the most important outstanding problems in modern physics.
Despite the overwhelming astrophysical evidence for its existence, no conclusive terrestrial observation of DM has yet been reported \cite{Liu2017,Bertone2018}.
Currently, most of the effort in the search for DM has focussed on weakly interacting massive particles (WIMPs) with masses $\gtrsim10-100\,$GeV through their hypothesised non-gravitational interactions with Standard Model particles.
In this work, we consider low-mass WIMP DM with masses on the order of $1\,$GeV.

One long standing claim of a potential DM detection was made by the DAMA Collaboration, which uses a NaI-based scintillation detector to search for possible DM interactions within the crystal in the underground laboratory at the Gran Sasso National Laboratory, INFN, Italy~\cite{Bernabei2013} (see also Refs.~\cite{Bernabei2008,Bernabei2012}).
The results from the combination of the DAMA/LIBRA and DAMA/NaI experiments indicated an annual modulation in the event rate at around 3 keV electron-equivalent energy deposition (with a low-energy threshold of $\sim2$ keV) with a 9.3$\sigma$ significance~\cite{Bernabei2013}.
The phase of this modulation agrees very well with the assumption that the signal is due to the scattering of WIMP DM present in the galactic halo.
An annual modulation in the observed WIMP scattering  event rate is expected due to the motion of the earth around the sun, which results in an annual variation of the DM flux through a detector (and the mean DM kinetic energy); see, e.g., Refs.~\cite{Freese2013,Lee2015}.

Despite the significant signal, there is strong doubt that the DAMA Collaboration result can be due to WIMPs, since it is seemingly in conflict with the null results of many other direct detection experiments, e.g., Refs.~\cite{Aprile2019a,Aprile2018,LUXjan2017,Ren2018,Agnes2018}.
There are also several works which offer explanations the DAMA result in terms of non-DM origins, e.g., in Ref.~\cite{Pradler2013}.
However, it is not always possible to compare different experiments in a model-independent way, meaning it is difficult to make general statements to this effect.

For example, one possibility that has been considered in the literature is that the DAMA modulation signal may be caused by WIMPs that scatter off the atomic electrons~\cite{Bernabei2008a,Foot2018,Savage2009}, as opposed to nuclear scattering as is assumed in typical experiments.
This is particularly applicable for lighter WIMPs ($\lesssim$\,$10$\,GeV), which will not leave appreciable nuclear recoils.
Most direct detection experiments try to reject pure electron scattering events, in order to perform nuclear recoil searches with as low as possible background.
Conversely, the DAMA experiment is sensitive to WIMPs which scatter off either electrons or nuclei,
potentially allowing electron-interacting DM to explain the DAMA modulation while avoiding the tight constraints from other experiments.
In a recent work~\cite{RobertsDAMA2016}, however, we used scintillation and ionisation signals from the XENON100~\cite{XENONcollab2015a} and XENON10~\cite{Angle2011} experiments to rule out this possibility for the observed signal above 2\,keV; see also Refs.~\cite{Xe100er2017,XENONcollab2015,Kopp2009}.

Recently, newer results from the DAMA/LIBRA-phase2 experiment have become available \cite{Bernabei2018} (see also \cite{Bernabei2017,Bernabei2015}).
These results strengthen the claim for a detected signal, with the significance of the annual modulation in the 2--6\,keV energy window rising to 12.9$\sigma$.
Importantly, the low-energy threshold has been lowered in the new experiment to 1\,keV, and the annual modulation is also clearly present in this region (9.5$\sigma$ significance).
This may be of particular significance for the interpretation in terms of electron interacting DM.
In our previous work~\cite{RobertsDAMA2016} we showed that there would be an almost exponential increase in the potential event rate at lower energies for such models of light ($\sim$\,1\,GeV) WIMPs.

For the $\sim$\,keV energy depositions of interest to this work, the relevant process for electron-scattering DM is atomic ionisation.
Such processes are kinematically disfavoured at these energy scales, and therefore the scattering probes deep inside the bound-state wave function, with the main contribution coming from wavefunction at distances much smaller than the characteristic Bohr radius of an atom.
In such a situation, incorrect small-distance scaling of the wavefunctions (for example, by using an ``effective $Z$'' model, or assuming plane waves for the outgoing ionisation electron) can lead to large errors in the predicted ionisation rates~\cite{RobertsDAMA2016}.
Further, the relativistic effects for the electron wavefunction are crucial and must be taken into account~\cite{RobertsAdiabatic2016}.
As such, interpretation in terms of light WIMPs requires non-trivial calculations of the atomic structure and ionisation rates.
Finally, we note that there are several ongoing experiments \cite{Akerib2018,Akerib2018-er,Akerib2017,XMASScollaboration2018,Ema2018,Cappiello2019a,Antonello2019,Collaboration2019a} 
and proposals \cite{Essig2017,Kouvaris2016,McCabe2017a,Wolf2018a,Emken2019,Bringmann2018,Kurinsky2019} to search for light WIMPs in direct detection experiments.
We also note that weak evidence for annual modulation at 2\,keV from the COSINE collaboration has been recently made public~\cite{Collaboration2019} (see also Ref.~\cite{TheCosine-Collaboration2018}).

\section{Theory}
\subsection{Atomic ionisation}

Throughout the text and in the figures, we use relativistic units ($\h=c=1$), with masses, energies and momenta presented in eV, as is standard in the field.
However, it is also customary, e.g., to present cross sections in $\un{cm}^2$, event rates in counts/kg/keV/day.
Further, for the calculations of atomic ionisation, it is convenient and common to use atomic units ($\h=m_e=1$, $c=1/\a$).
Therefore, to avoid any possible confusion, we leave all factors $\h$ and $c$ in the equations.

We consider DM particles that have electron interactions of the form
\be\label{eq:hint}
V(\v{r}) = \h c\a_\chi \frac{e^{-\mu r}}{{r}},
\ee
where $\mu$ is the inverse of the length-scale for the interaction, set by the mediator mass (e.g.,\ $\mu=m_v c/\h$), and $\alpha_\chi$ is the effective DM--electron coupling strength.
Such effective interaction Hamiltonians arise generally in the case of either scalar or vector interactions (e.g., via the exchange of a dark photon).
The coefficient in (\ref{eq:hint}) is chosen so that in the case of a massless mediator (long-range interaction, $\mu=0$), this reduces to a Coulomb(-like) potential (with $\a\to\a_\chi$).
In the limit of a very heavy mediator, the above reduces to the contact interaction:
$V(\v{r}) = 4\pi \h c ({\a_\chi}/{\mu^2})\delta(\v{r})$.

The differential cross section (for fixed velocity $v$) for the excitation of an electron in the initial state $njl$ is
\be\label{eq:dsigma}
\frac{\d\s_{n jl}}{\d E} = 8\pi\a_\chi^2  \left(\frac{c}{v}\right)^2 \int_{q_-}^{q_+}\frac{q\d q}{(q^2+\mu^2)^2}\, \frac{K_{n jl}(E,q)}{E_H},
\ee
where $\h q$ is the magnitude of the momentum transfer, $E$ is the energy deposition, and $K$ is the atomic excitation factor, defined \cite{RobertsAdiabatic2016}
\be\label{eq:Knk}
K_{n jl} \equiv E_H\sum_{m}\sum_{f}\left| \bra{f}e^{i\v{q}\cdot\v{r}}\ket{nj l m}\right|^2 \varrho_f(E).
\ee
Here, $\varrho_f$ is the density of final states, $m=j_z$, and the total cross section is to be summed over all electrons $\d\s = \sum\d\s_{n jl}$.
The factor of the Hartree energy unit ($E_H \equiv m_e c^2\a^2\simeq 27.2\un{eV}$) is included in Eqs.~(\ref{eq:dsigma}) and (\ref{eq:Knk}) in order to make the $K$ factor dimensionless ($q$ and $\mu$ have dimensions of inverse length).
Since we are considering ionisation processes, the final state is a continuum electron with energy $\varepsilon = E - I_{njl}$ ($I_{njl}$ is the ionisation energy).
Formulas for calculating  the atomic excitation factor (\ref{eq:Knk}) are given in Appendix~\ref{sec:appendix}.

Equation (\ref{eq:dsigma}) is to be integrated over all possible values for the momentum transfer.
From conservation of momentum, the allowed values fall in the range between
\be\label{eq:qpm}
\h q_{\pm} =  m_\chi v \pm \sqrt{ m_\chi^2 v^2-2m_\chi E},
\ee
where both the DM particle and ejected electron are assumed to be non-relativistic.
For $\sim$\,GeV WIMPs leaving $\sim$\,keV energy depositions, the typical momentum transfer is  $q\sim\sqrt{2m_\chi E}\sim{\rm MeV}$, which is very large on atomic scales \cite{RobertsDAMA2016}.

The resulting differential event rate (per unit mass of target material) is proportional to the cross section (\ref{eq:dsigma}) averaged over incident DM velocities:
\be
dR = \frac{n_T \rho_{\rm DM}}{m_\chi c^2} \frac{\d\braket{\s_{n jl}v_\chi}}{\d E} \, \d E,
\ee
where $n_T$ is the number density of target atoms (per unit mass), and $\rho_{\rm DM}\sim0.3-0.4\un{GeV}\un{cm}^{-3}$ is the local DM energy density~\cite{Bovy:2012tw}.
We follow Ref.~\cite{EssigPRL2012} and parameterise the velocity-averaged cross section as
\be
\frac{\braket{\d\s v}}{\d E}  = \frac{\bar\s_e }{2 m_e c}\int\d v \frac{f(v)}{v/c}\int\limits_{q_-}^{q_+} a_0^2q\d q \, |F^{\mu}_{\rm \chi}(q)|^2\,{K(E,q)},
\ee
where $\sbe$ is the free electron cross section at fixed momentum transfer of $q=a_0^{-1}$, 
$a_0 = \h/(m_e c\a)$ is the Bohr radius, 
$\a\approx1/137$ is the fine-structure constant,
and $f$ is the DM speed distribution (in the lab frame, normalised to $\int f(v)\,\d v=1$).
In the case of a vector or scalar mediated interaction such as (\ref{eq:hint}), these are expressed
\begin{align}
\sbe &= a_0^2 \frac{16\pi\,\a^2\a_\chi^2}{((m_v/m_e)^2+\a^2)^2} \\
F_\chi(q) &= \frac{(m_v/m_c)^2 + \a^2}{(m_v/m_c)^2+ (\a a_0 q)^2}.
\end{align}
We have assumed here that $m_\chi\gg m_e$, which is valid for the considered mass ranges.
In the limit of a heavy mediator (contact-like interaction), the DM form factor reduces simply to $F_\chi=1$, 
while for an ultra-light mediator (Coulomb-like interaction), it reduces to $F_\chi = (a_0q)^{-2}$.
This is a convenient way to parameterise the calculations, and allows for easy model independent comparison between different results.

There is no contribution to the event rate stemming from DM velocities below $v_\min = \sqrt{2E/m_\chi}$, the minimum required to deposit energy $E$.
If the majority of the target momentum-space wavefunction density lies inside the $(q_-,q_+)$ region, then the integration over $q$ in Eq.~(\ref{eq:dsigma}) is essentially independent of the integration limits, so one may write
\[
\braket{\d\s v}  \propto  \eta(v_{\rm min}) \int q\d q\,|F_{\rm DM}(q)|^2K(E,q) \, \d E,
\]
where $\eta(v_{\rm min})$ is the mean inverse speed of DM particles fast enough to cause the ionisation ($v>v_{\rm min}$) for the given velocity distribution.
This is a common way to calculate DM direct detection event rates, particularly for nuclear recoils, however, we note that in the case of electron scattering it is not valid.
The cross section depends strongly on $q_-$ and hence the DM velocity, since, in many cases, the bulk of the electron momentum-space wavefunction lies below the allowed region for momentum transfer.
This means that a careful treatment of the DM speed distribution, including uncertainties, is required for the analysis (see also Ref.~\cite{WuFreese2019}).

\subsection{Calculation of the atomic ionisation factor}

\begin{figure}
\includegraphics[width=0.48\textwidth]{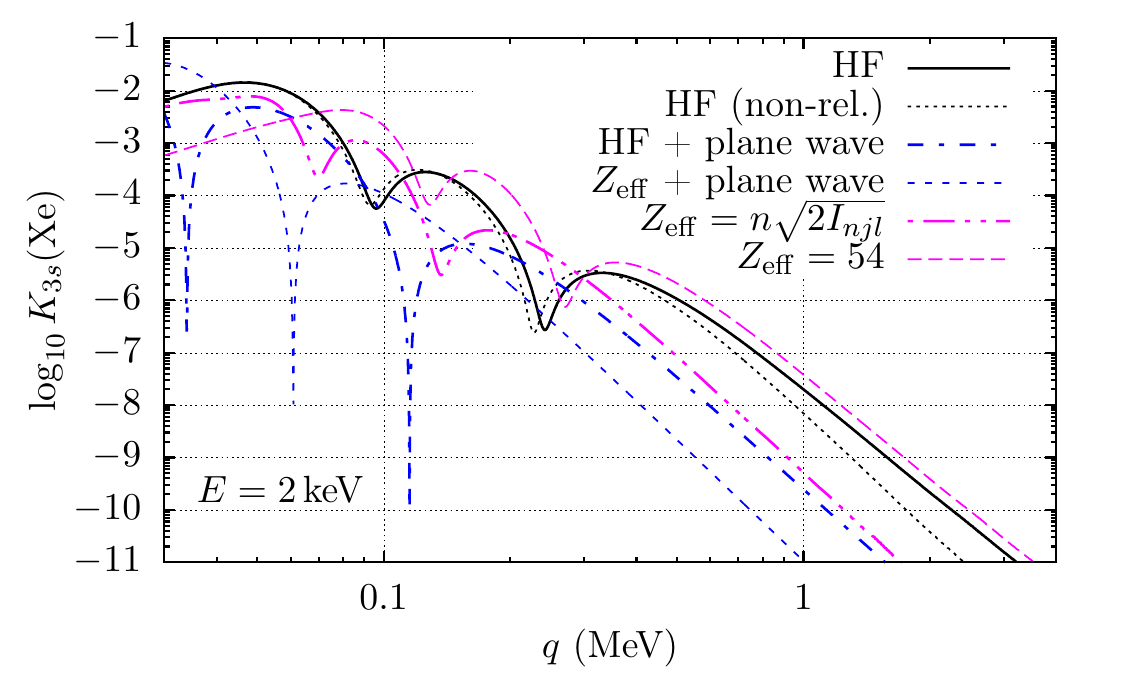}
\caption{Atomic factor $K$ (\ref{eq:Knk}) for the $3s$ state of Xe at fixed energy (2\,keV), calculated using different approximations.
For $3s$ Xe state, we have $Z_{\rm eff} = n\sqrt{2I_{n jl}/E_H}\simeq28$.
}
\label{fig:Kapprox}
\end{figure}

In Fig.~\ref{fig:Kapprox} we show a comparison of the atomic ionisation factor as calculated using a number of different approximations, as a function of the momentum transfer $q$ (for fixed $E$).
For the values relevant to this work, around $q\sim1\,$MeV, there is almost four orders of magnitude difference between the various approximations.
Also note that the relativistic effects are very important for large $q$, and the corrections continue grow with increasing $q$ \cite{RobertsAdiabatic2016}.

Since the typical kinetic energy of a $\sim\,$GeV mass WIMP is large compared to typical atomic transition energies, the minimum momentum transfer is given
$\h q_{\rm min} 
\sim m_\chi v_\chi \sim E/v_\chi$
(\ref{eq:qpm}).
Therefore, we see that 
$\h q_{\rm min}\gtrsim m_e v_e^2/v_\chi = p_e(v_e/v_\chi)\gg p_e$,
with $v_e\sim \a c - Z\a c$ the typical velocity of an atomic electron. 
The consequence is that only the very high-momentum tail of the wavefunctions (in momentum space) can contribute to such processes.
In position space, this part of the wavefunction comes from distances very close to the nucleus.
For a detailed discussion, see Ref.~\cite{RobertsAdiabatic2016}.

Therefore, care must be taken to perform the calculations of such processes correctly.
For example, it is common to calculate such processes using analytic hydrogen-like wavefunctions, with an effective nuclear charge, which is chosen to reproduce experimental binding energies: $Z_{\rm eff} = n\sqrt{2I_{n jl}/E_H}$.
While such functions give a reasonable approximation for low $q$, for the large $q$ values important for this work, they drastically underestimate the cross-section.
This is because such functions have incorrect scaling at distances close to the nucleus, which is the only part of the electron wavefunction that can contribute enough momentum transfer.

Another common approach is to approximate the outgoing ionisation electron wavefunction as a plane wave state.
Such functions also have the incorrect scaling at small distances, and underestimate the cross section by orders of magnitude for large $q$.
(This is mostly due to the missing Sommerfeld enhancement as discussed in Ref.~\cite{Essig2012}.)
More details regarding this point are given in Appendix~\ref{sec:appendix}.

Therefore, to perform accurate calculations one must employ a technique known to accurately reproduce the electron orbitals.
Namely, the relativistic Hartree-Fock method, including finite-nuclear size, and using continuum energy eigenstates as the outgoing electron orbitals.
Detailed calculations and discussion was presented in Ref.~\cite{RobertsDAMA2016}.
Formulas are given in the appendix  \ref{sec:appendix}.

\begin{figure*}
\includegraphics[width=0.47\textwidth]{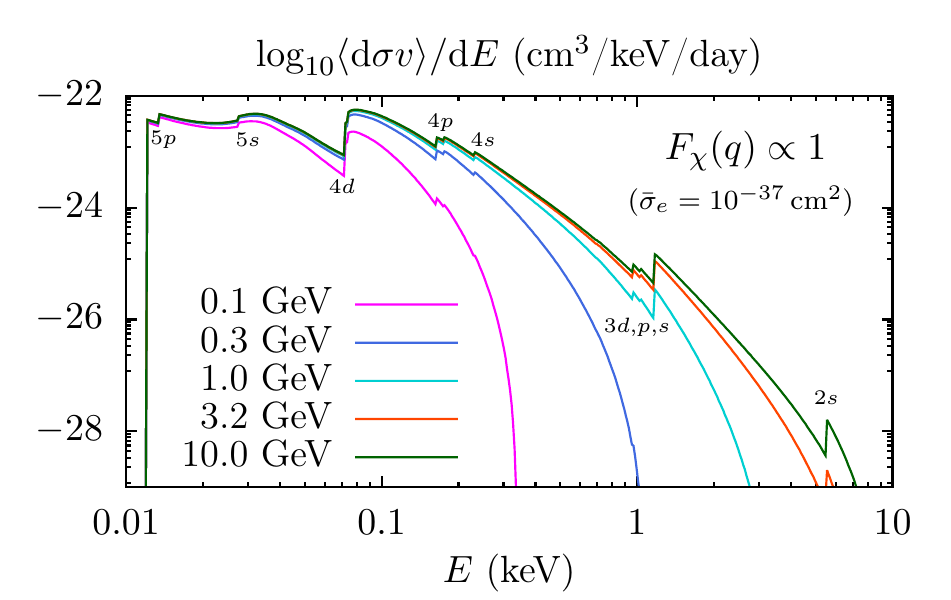}
\includegraphics[width=0.47\textwidth]{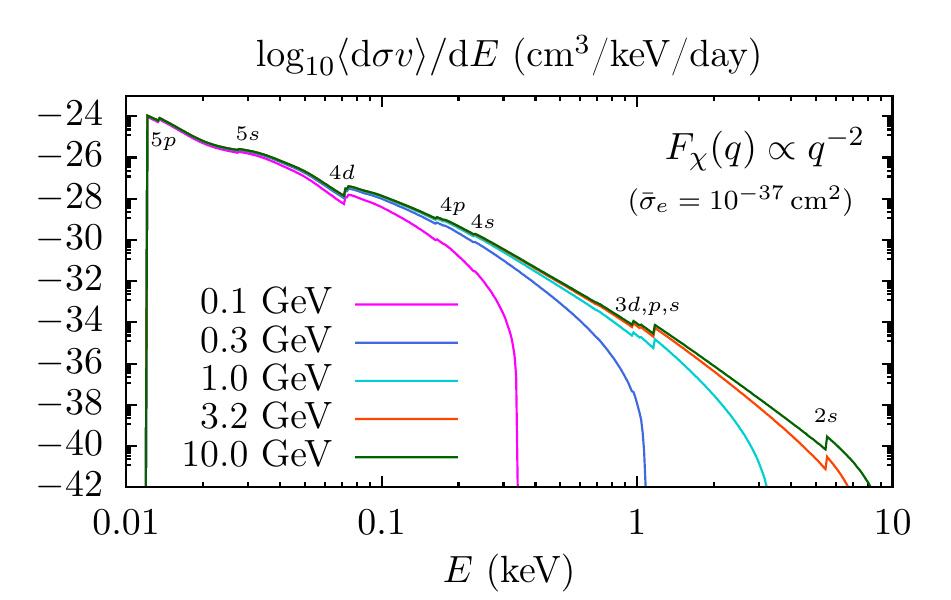}
\caption{
Velocity-averaged differential cross section for a single Xe atom, with $\sbe=10^{-37}\un{cm}^{2}$.
The left panel is for a contact (heavy mediator) interaction, and the right is for a long-range (Coulomb-like) interaction.
The kinks in the plots are due to the opening up of deeper electron shells.
There is no signal above $E_{\rm max}=m_\chi v_{\rm max}^2/2$, where $v_{\rm max}$ is the maximum DM velocity.
}
\label{fig:dsvde}
\end{figure*}

Given the extreme dependence on the atomic physics seen in Fig.~\ref{fig:Kapprox}, it is important to estimate the uncertainty in the calculations.
To gauge this, we also calculation the cross-section using other (simpler) methods.
Namely, we exclude the effect of the exchange potential from the Hartree-Fock method, and also solve the Dirac equations using only a local parametric potential (chosen to reproduce the ionisation energies) instead of the Hartree-Fock equations.
The effect this has on the calculations is very small, with the main difference coming from small changes in the calculated values for the ionisation energies.
This is as expected, since the cross-section is due mainly to the value of the wavefunctions on small distances, close to the nucleus, where many-body electron effects are less important (but the correct scaling is crucial).
All of these methods (unlike the effective $Z$ method, or plane-wave assumption) give the correct small-$r$ scaling of the bound and continuum electron orbitals.

The finite-nuclear size correction is important for large values of $q$, but is small compared to the relativistic corrections, and ultimately is not a leading source of error.
In any case, we include this in an {\em ab initio} manner, by directly solving the electron Dirac equation in the field created by the nuclear charge density, which we take to be given by a Fermi distribution,
$\rho(r) = \rho_0 [1+\exp(r-c/a)]^{-1}$.
Here $t\equiv 4a\ln3\simeq2.3\,{\rm fm}$ and $c\simeq 1.1 A^{1/3}\,{\rm fm}$ are the nuclear skin-thickness and half-density radius, respectively, e.g.,~\cite{Fricke1995}, and $\rho_0$ is the normalisation factor.
We note that the uncertainties stemming from the atomic physics errors are small compared to those coming from the assumed dark matter velocity distribution and detector performance, as discussed in the following sections.

Plots of the velocity averaged differential cross-sections for several WIMP masses and mediator types are presented in Fig.~\ref{fig:dsvde}.
We find very good agreement with similar recent calculations for Xe atoms in Ref.~\cite{Pandey2018}.
We present these plots for the xenon atom, since it is the most common target material.
For DAMA/LIBRA experiment, the cross section is dominated by scattering off iodine ($Z=53$), which has very similar electron structure to xenon ($Z=54$).

\subsection{Annual modulation}

We assume the DM velocity distribution is described by the standard halo model, with a cut-off (in the galactic rest frame) of
$v_{\rm esc} = 550(55)\un{km/s}$, and a circular velocity of $v_0=220(20)\un{km/s}$, see, e.g., Ref.~\cite{Freese2013,Baum2018}.
The numbers in the parentheses above represent estimates for the uncertainties in the values.
This is important, due to the strong velocity dependence of the cross section (see also, e.g., Ref.~\cite{WuFreese2019}).
We use these uncertainties to estimate the resulting uncertainty in the calculated event rates.

For the calculations, the velocity distribution is boosted into the Earth frame, which has a speed of
\be v_E(t) \approx v_L + v_{\rm orb}\cos\beta\cos({2\pi}\cdot{{\rm yr}^{-1}} \, t + \phi). \ee
Here, $v_L = v_0+13\un{km/s}$ is the average local rest frame velocity, accounting for the peculiar motion of the Sun, 
$v_{\rm orb}$ is the Earth's orbital velocity, 
and $\cos\beta\approx 0.49$ accounts for the inclination of Earth's orbit to the galactic plane.
The $\cos(\w t)$ term accounts of the annual change in the local frame velocity due to the orbital motion around the sun,
with phase $\phi$ chosen such that $v_E$ is maximum at June 2, when the Earth and sun velocities add maximally in the galactic halo frame.

Due to the strong velocity dependence of the cross-section, the resulting event rates are not perfectly sinusoidal,
particularly at higher energies and lower WIMP masses \cite{RobertsDAMA2016}.
However, the general sinusoidal feature remains a reasonable approximation.
We define the modulation amplitude as $(R_{\rm max} - R_{\rm min})/2$.

\section{Implications from DAMA/LIBRA-phase2}

\begin{figure*}
\includegraphics[width=0.47\textwidth]{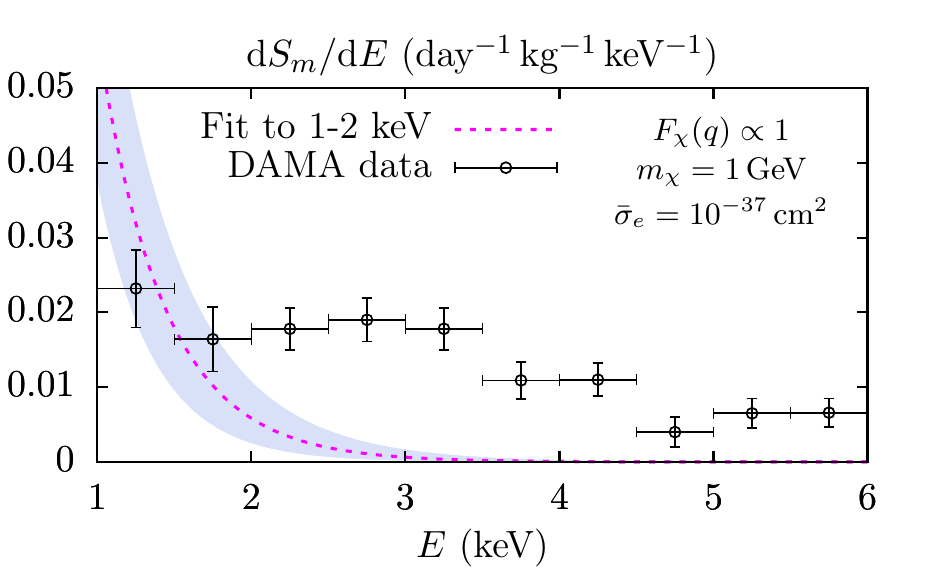}
\includegraphics[width=0.47\textwidth]{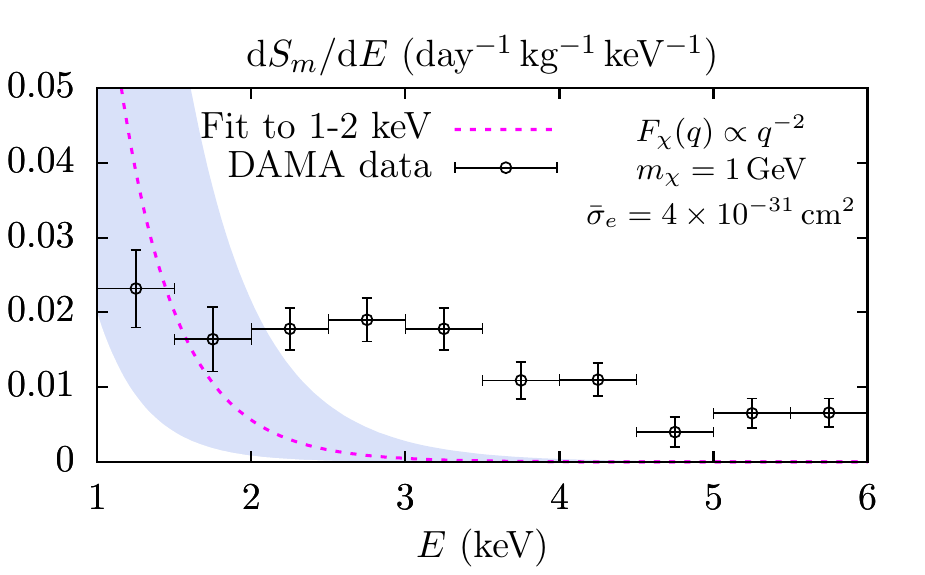}
\caption{
Calculated modulation amplitude for NaI, accounting for the DAMA detector resolution, with $\sbe$ chosen to reproduce the observed DAMA/LIBRA-phase2 modulation amplitude averaged across the 1--2\,keV energy bin.
The black points are the combined DAMA/LIBRA data (the points below 2\,keV from phase2 alone) \cite{Bernabei2018}.
The shaded blue region shows the uncertainties from the calculations (this work), which are mostly due to uncertainties in the DAMA energy resolution.
The plot is drawn assuming $m_\chi=1$\,GeV for a contact-like interaction (left), and a long-range interaction (right).
Clearly, the fit is poor.
}
\label{fig:fitSpectrum}
\end{figure*}

In order to calculate the number of events detected within a particular energy range, the energy resolution of the detectors must be taken into account. 
To do this, we follow the procedure from Ref.~\cite{Bernabei2008a}, and take the detector resolution to be described by a Gaussian with standard deviation
\be
\sigma_{\rm LE}/E = \a_{\rm LE}/\sqrt{E/{\rm keV}} + \b_{\rm LE},
\ee
which is measured at low energy to be given by $\a_{\rm LE} = 0.45(4)$, and $\b_{\rm LE}=9(5)\E{-3}$ \cite{Bernabei2008b}.
The calculated rate, $R$, is integrated with the Gaussian profile to determine the observable event rate, $S$:
\be
\frac{\d S}{\d E} = \int_{E_{\rm HW}}^\infty g_{\sigma_{\rm LE}}(E'-E) \, \frac{\d R(E')}{\d E'} \, \d E',
\ee
where $E_{\rm HW}$ is the hardware threshold, which for DAMA is 1 photoelectron \cite{Bernabei2008b}.
The extracted number of photoelectrons is measured by the DAMA collaboration to be 5.5--7.5 photoelectrons/keV, depending on the detector \cite{Bernabei2008b}.
We take an average value of 6.5, with $\pm1$ as an error term, so that $E_{\rm HW} = 0.15(3)\un{keV}$.
We don't take the detector efficiency into account, because the DAMA collaboration present their results corrected for this~\cite{Bernabei2018}.

The effect of the finite detector resolution is that it allows events that originally occur at lower energies to be visible in the observed data above the threshold.
This is particularly important due to the strong enhancement in the cross section at low energies (see Fig.~\ref{fig:dsvde}).

Due to the strong atomic number $Z$ dependence, the cross section for scattering off sodium electrons is negligible \cite{RobertsDAMA2016}.
So, for the DAMA NaI crystals, it is sufficient to calculate the rate due to scattering of the iodine electrons.
We have treated iodine as though it were a free atom, whereas in fact, it is bound in the NaI solid.
Only the outermost $5p$ orbitals are involved in binding.
However, even after accounting for the detector resolution, the $5p$ (and $5s$) orbitals contribute negligibly, with the dominant contribution at $\sim$\,1--2\,keV coming from the inner $3s$ shell, which is very well described by atomic wavefunctions.

Using this approach, we calculate the expected event rate and annual modulation amplitude for DAMA, as a function of the incident WIMP mass, assuming both an ultra-light and super-heavy mediator.
Due to the very large enhancement in the expected event rate at smaller energies, the calculated modulation amplitude is a poor fit to the observed DAMA spectrum.
In Fig.~\ref{fig:fitSpectrum}, we present the calculated spectrum along-side the DAMA/LIBRA-phase2 data~\cite{Bernabei2018}.
For the coupling strength (parameterised in terms of $\sbe$), we have fitted the expected event rate to the observed DAMA modulation signal only for the lowest $1-2$\,keV bins. Taking the higher energy bins into account can only increase the best-fit value for $\sbe$, so (as discussed below) this is the most conservative choice.

\begin{figure*}
\includegraphics[width=0.47\textwidth]{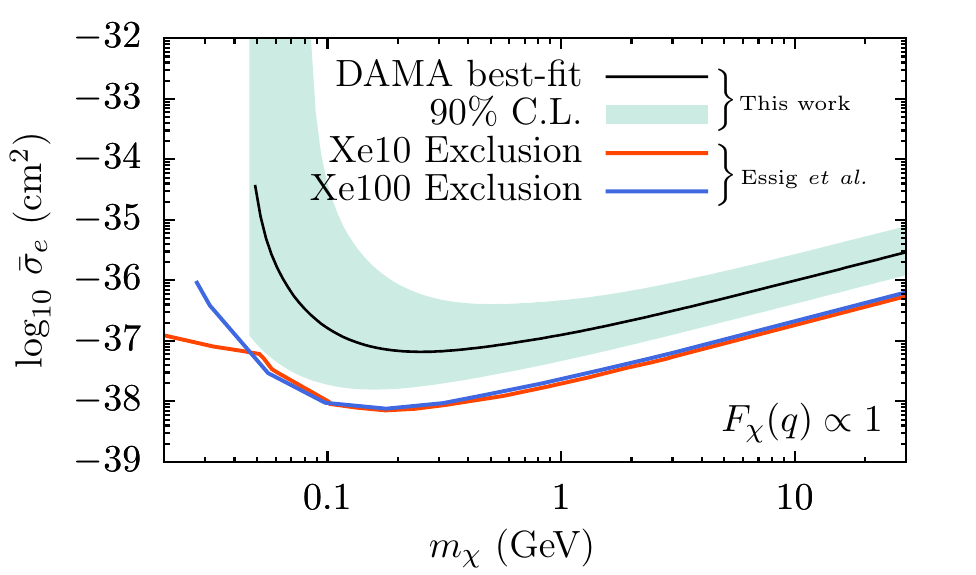}
\includegraphics[width=0.47\textwidth]{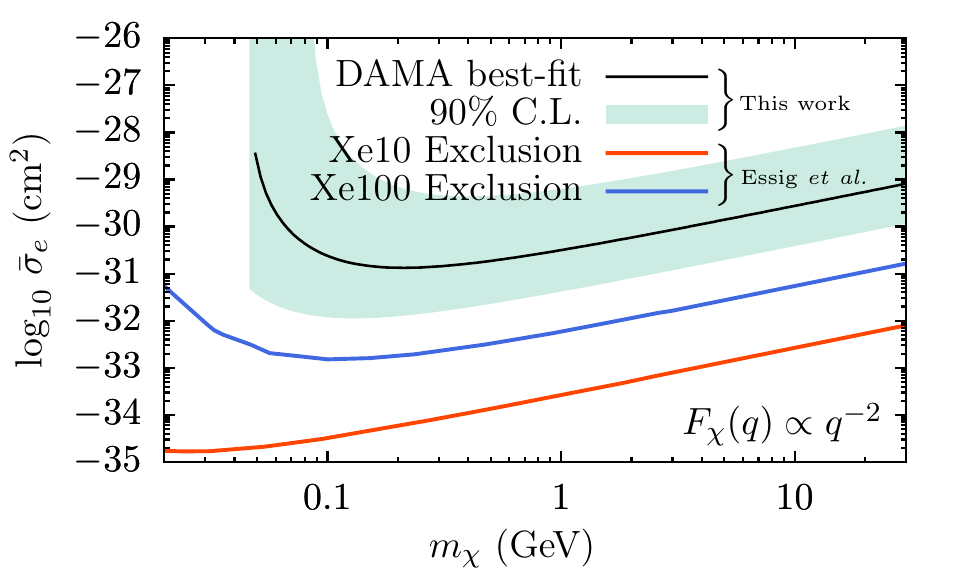}
\caption{
The black line shows the calculated value for $\sbe$ required to reproduce the observed DAMA modulation signal over the 1--2\,keV energy bin, as a function of the WIMP mass for the heavy (left) and ultra-light (right) mediator cases.
The green shaded region shows the 90\% C.L. region for the fit, taking into account uncertainties stemming from the detector resolution, standard halo model, and atomic physics errors.
The red and blue curves are the 90\% C.L. exclusions from Ref.~\cite{Essig2017} (Essig {\em et al.}), derived from the $S2$ ionization signals from the XENON10 and XENON100 experiments, respectively.
The ``DAMA-allowed'' regions are excluded for all relevant WIMP masses by these bounds.
The fit for DAMA was performed by averaging over just the lowest energy bins without regard to the shape of the spectrum.
Taking the higher energy bins into account pushes the DAMA region higher, strengthening this conclusion (see Fig.~\ref{fig:fitSpectrum}).
}
\label{fig:Constraints}
\end{figure*}

In Fig.~\ref{fig:Constraints} we plot the best-fit regions for the lowest energy DAMA/LIBRA-phase2 modulation signal, as a function of possible DM masses $m_\chi$ and coupling strengths $\sbe$.
Despite the large enhancement in the expected event rate at the lower energies, and the conservative assumptions made for extracting the best-fit, the interpretation of the observed modulation amplitude in terms of electron-interacting dark matter is inconsistent with existing bounds.
All regions of parameter space that could possibly explain the observed DAMA signal are excluded by constraints derived in Ref.~\cite{Essig2017}, using $S2$ ``ionisation-only'' results from the XENON10~\cite{Angle2011} and XENON100~\cite{XENON100-LowM-2016} experiments.

Note the large uncertainties visible in the plots in Figs.~\ref{fig:fitSpectrum} and \ref{fig:Constraints}.
The dominating source of error comes from the uncertainties in the detector response and energy resolution.
Sizable errors also arise due to uncertainties in the standard halo model DM velocity distribution.
Uncertainties coming from the atomic physics calculations are also included, but are negligible.
The uncertainties in the detector resolution and DM velocities themselves are not so large ($\sim$\,10\%) -- but they lead to very large uncertainties (up to an order of magnitude) in the observable event rate. 
This is due to the very strong enhancement in the event rate at low energies, which makes the observed rate very sensitive to the detector cut-offs and energy resolution.
Clearly, taking these uncertainties into account is crucial.

\section{Propsects for liquid xenon detectors}

In this section, we discuss the prospects for the detection of light ($\sim$\,GeV) WIMPs using xenon duel-phase time projection chambers.
(We base our discussion here on XENON Collaboration detectors, see e.g., Ref.~\cite{Aprile2012b}; similar principals apply for other experiments.)
When a scattering event occurs in the liquid xenon bulk of such a detector, a prompt $S1$ scintillation signal is induced, which is proportional to the total energy deposited in the detector.
Then, any ionized electrons are drifted upwards through the liquid/gas boundary (via an applied electric field), where a secondary scintillation signal ($S2$) that is proportional to the number of ionized electrons may be observed.
Combining the $x,y$ spatial resolution of the top and bottom photodetectors with the $z$-resolution from the time between the $S1$ and $S2$ signals allows three-dimensional reconstruction of the event geometry.
This allows for the ``fiducialization'' of the target material, where only scattering events occurring within the inner volume of the detector are included in the analysis.
This is an important stage of background rejection, since charged particles are much more likely to scatter quickly, i.e.,~at the outer regions of the xenon chamber, whereas feebly interacting particles such as WIMPs are equally likely to scatter anywhere within the detector volume.
Further, the ratio between the relative strengths of the $S1$ and $S2$ signals may be used to distinguish between nuclear and electronic scattering events.
The combination of both the $S1$ and $S2$ signals is thus key to understanding the source of any scattering events.

Proposals to use the ionization-only ($S2$) signals to search for sub-GeV WIMPs have been made previously~\cite{Essig2012,EssigPRL2012}, and limits from $S2$ observations using XENON10 and XENON100 experiments have been set\cite{Essig2017} (as discussed in the previous section).
It is worth noting that the best constraints actually come from the older XENON10 experiment (finished in year 2011), despite its much smaller detector mass, older generation of detectors, and much smaller total exposure.
This is due to the detection strategy of the modern experiments, which rely on the combination of $S1$ and $S2$ signals.

The reason $S2$ ionisation-only signals were considered, is because for low-mass WIMPs, the typical energy deposited in the detectors is much smaller than the $\sim$\,few keV effective low-energy threshold for $S1$ signals.
It was thus believed that the $S1$ scintillation signal produced from such events would be negligible.
In this work, we demonstrate that, due to the large enhancement from lower energies and the finite detector resolution, the prompt $S1$ scintillation signal can be many times larger than otherwise expected, and that it therefore can be a promising WIMP direct detection observable.
Thus, it would be possible to perform a low-mass WIMP search with modern liquid xenon detectors using the combined $S1$ and $S2$ signals.
Detailed calculations of the observable $S2$ spectrum from low mass WIMPs was presented recently in Ref.~\cite{Essig2017}; here we present calculations for the corresponding $S1$ signal.

\begin{figure*}
\includegraphics[width=0.47\textwidth]{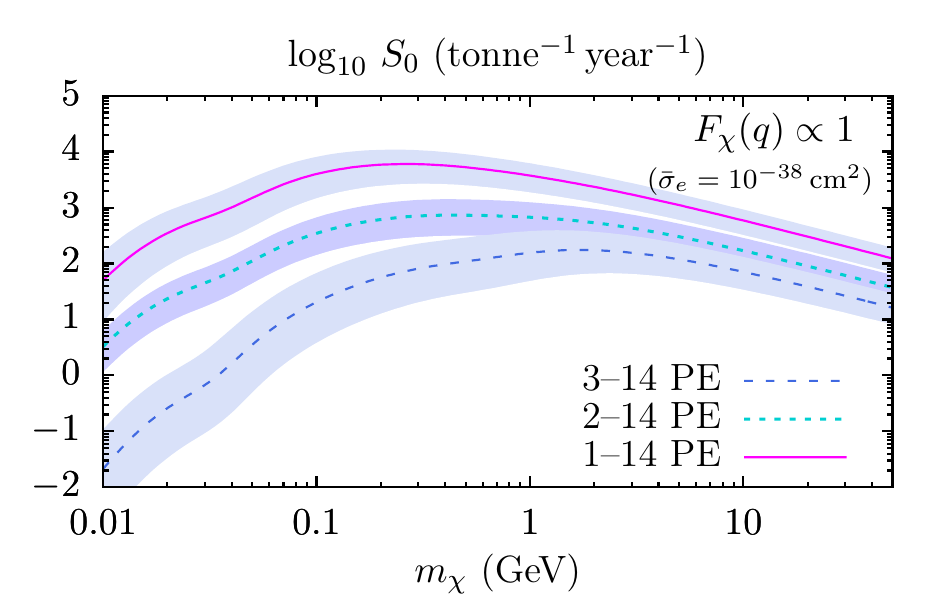}
\includegraphics[width=0.47\textwidth]{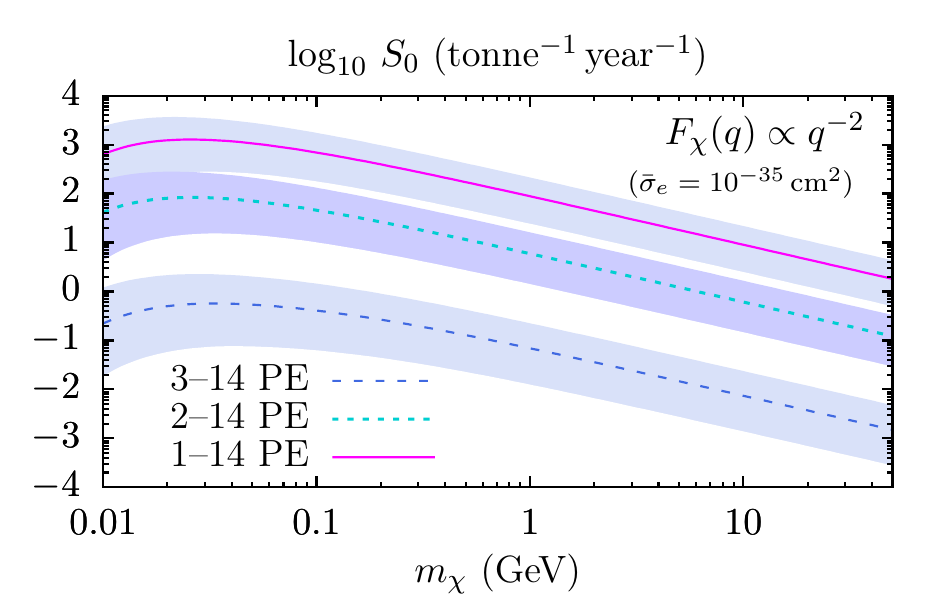}
\caption{Hypothetically observable WIMP electron recoil event count expected for a 1 tonne$\cdot$year exposure of a liquid xenon detector (based on XENON100) using the prompt scintillation ($S1$) signal; ({\em left}) for a contact interaction (with $\sbe=10^{-38}\un{cm}^{-1}$), ({\em right}) for a contact interaction (with $\sbe=4\times10^{-35}\un{cm}^{-1}$).
The shaded blue regions show the uncertainties in the calculations.
The $\sbe$ values are chosen to be below the present constraints, which are most stringent around $\sim0.1\un{GeV}$ (see Fig.~\ref{fig:Constraints}).
Note that for larger masses, the constraints are less stringent, and larger events rates are not ruled out.
}
\label{fig:Xe100-fit}
\end{figure*}

\begin{figure}
\includegraphics[width=0.48\textwidth]{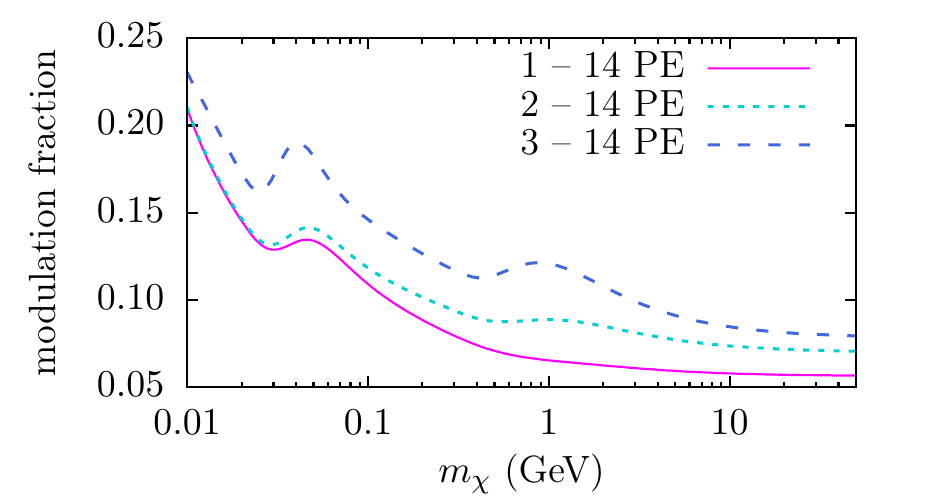}
\caption{
Expected fractional modulation amplitude
[$(S^{\rm max} - S^{\rm min})/2S^{\rm avg}$] 
as a function of the WIMP mass, assuming a heavy mediator ($F_\chi=1$),
for the prompt $S1$ signal in a XENON100-like detector, in the 1--, 2--, and 3--14 PE bins.
A discussion of the features is given in the text.
}
\label{fig:modFrac}
\end{figure}

We calculate the potentially observable $S1$ (prompt scintillation signal) event rate and modulation amplitude for a hypothetical future liquid xenon detector.
We model this detector after that of XENON100, and follow Ref.~\cite{AprileAxion2014} for the conversion from the energy deposition to the observable photoelectron (PE) count  (see also Ref.~\cite{AprileAP2014}).
In this case, the relevant quantity is a counted rate as a function of observable photoelectrons, denoted $s1$.

The calculated event rate for the production of $n$ photoelectrons is obtained by applying Poisson smearing to the calculated differential rate \cite{AprileAxion2014}.
We do this according to a Poisson distribution,
$P_n(N) = e^{-N}({N^n}/{n!})$,
where $N=N(E)$ is the expected/average number of photo-electrons produced for a given energy deposition $E$, and $n$ is the actual number of photoelectrons produced.
The relation between the deposited energy (electron recoil energy) and the produced number of photoelectrons is given in Fig.~2 of Ref.~\cite{AprileAxion2014}. 
We model this as a power law: $N(E) = a E^b$, with $a=1.00(25)$ and $a=1.53(10)$, which give the best fit for the lower energies applicable for this work, accounting for the uncertainties from Ref.~\cite{AprileAxion2014}.

Further, to account for the photomultiplier tube (PMT) detector resolution, we convolve the calculated rate with a Gaussian of standard deviation $\sigma = \sigma_{\rm PMT}\sqrt{n}$, with $\sigma_{\rm PMT}=0.5\,$PE \cite{AprileAP2014}.
We do not include uncertainty contributions from the PMT resolution, though note that we have checked, and error in the $N(E)$ conversion is by far the dominant source of uncertainty in this step.
Finally, the detection acceptance is taken into account as $\epsilon(s1) = 0.88(1-e^{s1/3})$ \cite{AprileAxion2014}, though we note that this has an insignificant impact on the results.
The final expression for the observable event rate, $S$, as a function of counted PEs $s1$ is
\be
\frac{\d S}{\d s1} =  \epsilon(s1) \sum_n g_\sigma(n-s1)\int_0^\infty \frac{\d R(E)}{\d E} P_n(N) \, \d E.
\ee

We calculate potential event rates, assuming a value for $\sbe$ that is not excluded by current experiment, for a one tonne-year exposure in Fig.~\ref{fig:Xe100-fit} as a function of the WIMP mass, for both a contact and long range interaction.
We show the rate integrated between $3$ and 14 PE, as in Ref.~\cite{XENONcollab2015} (see also Refs.~\cite{Xe100er2017,XENONcollab2015a}).
This roughly corresponds to the $2-6$\,keV energy window.
The rate is strongly dominated by the lower PE contribution, so it doesn't matter where the higher PE cut is taken.
We also present the expected rates for the ranges including 1 and 2 PE.
The larger and less-well understood background at these lower energies make experiments more difficult to interpret.
However, the much enhanced event rate, and large annual modulation amplitudes, may make these regions interesting for future experiments.

In Fig.~\ref{fig:modFrac}, we show the expected annual modulation fraction for the same type of experiment.
Due to the strong velocity dependence of the cross-section, the fractional modulation amplitude is large.
For example, for a $\sim$\,$0.1\,$GeV WIMP, where the event rate may be expected to be high, it is $\sim$\,15-20\%.
The peaks in the annual modulation curves (Fig.~\ref{fig:modFrac}) at around 0.04 and 1 GeV are due to the opening of the $n=3$ and $n=4$ shells in Xe.
Electrons may only become ionised if their binding energies are lower than the maximum kinetic energy of the incident WIMPs:
\be
I_{njl}<K_{\rm max} \simeq \frac{1}{2}m_\chi v_{\rm max}^2 \sim 4 \left(\frac{m_\chi c^2}{{\rm GeV}}\right) {\rm keV}.
\ee
The ionisation rate for shells with energies close to this number (i.e.,\ that are ``only just'' accessible) will be sensitive to small changes in the velocity distribution.
For Xe, these occur for the $n=3$ shell just above $\sim$\,$1\un{keV}$, and the $n=4$ shell just below $\sim$\,$0.1\un{keV}$ (see Fig.~\ref{fig:dsvde}).

\section{Conclusion}

We have calculated the expected event rate for atomic ionization by $\sim$\,GeV scale WIMPs that scatter off atomic electrons, relevant to the DAMA/LIBRA direct detection experiment.
Though the calculated event rate and annual modulation amplitude is much larger than may be expected, we show that such WIMP models cannot explain the observed DAMA modulation signal without conflicting with existing bounds, even when just the lowest energy $1-2$\,keV bin is fitted.
Taking higher bins into account strengthens this conclusion.
Further, we demonstrate explicitly the importance of treating the electron wavefunctions correctly, and note that the expected event rates are extremely sensitive to the detector resolution and low-energy performance, and the assumed dark matter velocity distribution.
Uncertainties in these quantities lead to large uncertainties in the calculated rates, and therefore must be taken into account.
Finally, we calculate the potentially observable event rate for the prompt scintillation signal of future liquid xenon detectors.
Large event rates would be expected for dark matter parameters which are not excluded by current experimental bounds, making this an important avenue for potential future discovery.

\acknowledgements
BMR gratefully acknowledges financial support from Labex FIRST-TF.
This work was also supported by the Australian Research Council and the Gutenberg Research College fellowship.

\appendix
\section{Atomic ionisation factor}\label{sec:appendix}

\subsection{Continuum final states (energy eigenstates)}

For the electron wavefunctions, we employ the Dirac basis, in which single-particle orbitals are expressed
\be\label{eq:psi-nkm}
\braket{\v{r}|nj l m} = \frac{1}{r}\twocomp{f_{n  jl}(r)\,\Omega_{j l m}(\vhat{n})}{i g_{n  jl}(r)\,\Omega_{j \tilde{l} m}(\vhat{n})},
\ee
where $\Omega$ is a spherical spinor,
\be
\Omega_{j l m} \equiv \sum_{\s} \braket{l,m-\s ,\,1/2,\s|j,m}\,Y_{l,m-\s}(\vhat{n})\,\chi_{\s},
\ee
with $\braket{j_1 m_1 j_2 m_2|JM}$ a Clebsch-Gordon coefficient, $Y_{lm}$ a spherical harmonic, 
$\tilde{l}=l\pm1$ for $j=l\pm{1}/{2}$,
and $\chi_{\s}$ a spin eigenstate with $\s=\pm{1}/{2}$ being the spin orientation. 
Note, that in the non-relativistic limit, the small component $g\to0$, and $f\to P$, where $P/r$ is the radial solution to the non-relativistic  Schr\"odinger equation.
To reach non-relativistic limit in the calculations, we allow the speed of light $c\to\infty$ inside the code before the Dirac equation is solved.
We also note that the relativistic enhancement discussed in the main text does not stem from the lower $g$ component, whose contributions to $R$ scale as $(Z\a)^2$ (except in the case of pseudo-scalar/pseudo-vector interactions, where the $g$ functions contribute at leading order \cite{RobertsDAMA2016}).
Instead, they come from differences in the radial dependence of the upper $f$ component \cite{RobertsAdiabatic2016}.

\begin{figure*}
\includegraphics[width=0.48\textwidth]{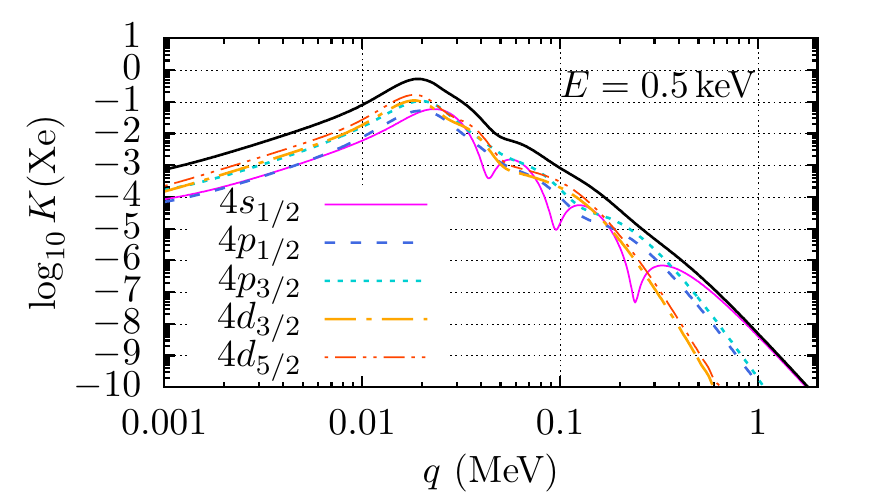}
\includegraphics[width=0.48\textwidth]{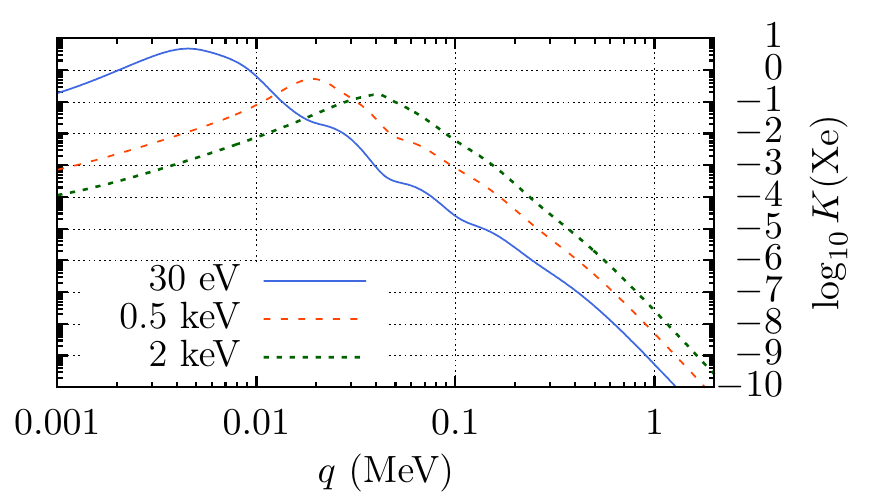}
\caption{
The left panel shows the dominating contributions to the atomic ionisation factor $K$ for $E=0.5\un{keV}$.
For a given energy, $K$ is dominated by the deepest accessible shell (lowest principal quantum number $n$).
The solid black line is the total sum (including states with higher $n$ that are not shown explicitly).
The main contribution at low $q$ comes from the states with highest total angular momentum $j$, while the main large $q$ contribution comes from states with the lowest orbital momentum $l$.
The right panel shows the total $K$ (summed over all accessible atomic electrons) for $E=0.03,\,0.5$, and 2\,keV, which are dominated by the $n=5$, $4$, and $3$ shells, respectively.}
\label{fig:Knk}
\end{figure*}

The continuum state orbitals are defined similarly,
\be\label{eq:psi-enkm}
\braket{\v{r}|\en j l m} = \frac{1}{r}\twocomp{f_{\en  jl}(r)\,\Omega_{j l m}(\vhat{n})}{i g_{\en  jl}(r)\,\Omega_{j \tilde{l}  m}(\vhat{n})}
\ee
with energy normalisation~\cite{BetheBook}, so that:
\be
\int_{\en-\delta\en}^{\en+\delta\en}\braket{\en'j l m|\en j l m}\,\d \en' = 1.
\ee
In practice, the normalisation is achieved by a comparison with analytic Coulomb functions at large $r$ \cite{BetheBook}.
Note, this formalism means that $\varrho_f$ (\ref{eq:Knk}) is included already in the definition of the orbitals $\braket{\v{r}|\en j l m}$, which have dimension $[{\rm Length}]^{-3/2}[{\rm Energy}]^{-1/2}$.

To calculate the atomic ionisation factor,
we first expand the exponential in Eq.~(\ref{eq:Knk}) as a sum over irreducible spherical tensors:
$e^{i\v{q}\cdot\v{r}}=\sum_{L=0}^\infty\sum_{M=-L}^{L}T_{LM}$ (see, e.g.,~\cite{Varshalovich1988}).
Then, from the standard rules for angular momentum, the atomic factor can be expressed as
\be
 K_{n jl}(E,q) = E_H\sum_L\sum_{ j'l'} 
|R_{n jl}^{j'l'L}(\en , q)|^2  C_{jl}^{j'l'L}.
\ee
Here, $R$ is the radial integral
\begin{multline}
R_{n jl}^{j'l'L}(\en , q)
 = \int_0^\infty \big[ f_{n jl}(r)f_{\en j'l'}(r) \\ + g_{n jl}(r)g_{\en j'l'}(r)\big]j_L(qr)\,\d r,
\end{multline}
where $\en = I_{n jl}-E$, $j_l$ is a spherical Bessel function, and
$C$ is an angular coefficient given by (for closed shells)
\be
 C_{jl}^{j'l'L} =
 [j][j'][L]\threej{j}{j'}{L}{-1/2}{1/2}{0}^2 \, \Pi_{ll'}^L, 
\ee
with $[J]\equiv2J+1$,
$\left(\ldots\right)$ being a Wigner $3j$ symbol,
 and $\Pi_{ll'}^L=1$ if $l+l'+L$ is even and is 0 otherwise.
The primed quantities refer to the angular momentum state of the ejected ionisation electron (final state).
For $q\gtrsim1\un{MeV}$, only the $L=0$ term contributes significantly, while for $q\lesssim0.01\un{MeV}$, only the $L=1$ term is important.
For the intermediate region $\sim$\,$0.1\un{MeV}$, the sum saturates reasonably rapidly, and convergence is reached by $L=4$.
These equations are valid for the case of DM that interacts via vector and scalar mediators; similar expressions for the case of pseudo-vector and pseudo-scalar mediators are given in Ref.~\cite{RobertsDAMA2016}.

Plots of the atomic ionisation factors showing the energy and momentum-transfer dependence, as well as the contributions from different atomic orbitals, are shown in Fig.~\ref{fig:Knk}.

\subsection{Plane wave final states}

Here we present the formulas for calculating the ionisation assuming a plane-wave final state.
This is done only as a demonstration; we stress, as discussed above, that this is not a reasonable approximation for the processes considered in this work.
Take the final ionisation electron state as a plane wave
\be\label{eq:psi-pw}
\braket{\v{r}|\v{p}} = e^{i\v{p}\cdot\v{r}/\h},
\ee
with $|p|=\sqrt{2m_e \en}$, subject to the normalisation
\[
\int\frac{\d^3\v{p}}{(2\pi \h)^3}\braket{\v{p}|\v{p}} = 1.
\]
The relativistic corrections to (\ref{eq:psi-pw}) are suppressed as $\sqrt{\en/(2m_ec^2)}$, and can be safely excluded.
If $\h\v{k}$ ($\h\v{k'}$) is the initial (final) WIMP momentum, the minimum allowable momentum transfer can be expressed $\h q_{-} = \h|\v{k}'-\v{k}|\approx E/ v \gtrsim m_ev_e^2/ v = p_e (v_e/v)$. 
Since for inner-shell electrons $v_e\sim Z\a c$, while the DM speed $v\sim10^{-3}c$, this implies $\h q\gg p_e$.
Then, the form factor can be expressed as
\be
K_{n jl}^{\rm pw} = \frac{2E_H}{\pi}|\Phi_{n jl}(q)|^2 \sqrt{{\frac{2 m_e^3(E-I_{n jl})}{\h^3}}}(2j+1), 
\ee
with the radial integral defined
\be
\Phi_{n jl}(q) = \int_0^\infty f_{n jl}(r)j_l(qr)\,r\,\d r.
\ee
This method of calculating the event rate is used widely in the literature, however, for large values of $q$ it can drastically underestimate the cross section by orders of magnitude (see main text).
This is due partly to the missing Sommerfeld enhancement, which was also discussed in the context of DM-induced ionisations in Ref.~\cite{Essig2012}.
The effect arises due to the attractive potential of the nucleus, which enhances the value of the unbound electron wavefunction near the nucleus.
As discussed above in the main text, it is only the portion of the wavefunction close to the nucleus that contributes to the cross section.

The size of the Sommerfeld enhancement can be estimated for hydrogen-like $s$-states as (in atomic units)
\be\label{eq:Fermi}
\frac{K_{ns_{1/2}}}{K^{\rm pw}_{ns_{1/2}}}\Bigg|_{r\to0} \approx \frac{8\pi Z}{\left[1-\exp(-\frac{2\pi Z}{|p'|})\right]n^3|p'|},
\ee
where each of the $K$ terms is calculated using only the leading small-$r$ terms in the expansion of the wavefunction, and $p'=\sqrt{2m_e\en}$ is the momentum of the outgoing electron.
In the non-relativistic limit, the contribution to $K$ coming from the first-order term in the small-$r$ expansion of the electron wavefunctions is identically zero, and the leading non-zero contribution only arises at second order.
In contrast, using relativistic functions, one finds that the lowest-order term survives, leading to significant relative enhancement due to relativistic electron effects \cite{RobertsAdiabatic2016}.
Therefore, the non-relativistic equation (\ref{eq:Fermi}) also underestimates the relative enhancement.
Scaling of inner-shell electron wavefunctions near the nucleus goes as $|\psi(0)|^2\sim Z^3$, as for hydrogen-like functions.
However, the scaling for outer-shell electrons is not as simple~\cite{LLVol3}, so it's important to use electron wavefunctions that correctly reproduce the low-$r$ behaviour, including the correct screening and electron relativistic effects (e.g., the relativistic Hartree-Fock method, as employed here).

\bibliography{DMe}

\end{document}